\documentclass[
pra,twocolumn,preprintnumbers,amsmath,amssymb]{revtex4}

\usepackage{graphicx}
\usepackage{amssymb}
\usepackage{amsmath}
\usepackage{dsfont}
\usepackage{bm}
\usepackage{mathrsfs}
\usepackage{times}

\begin{document}
\title{Making geometrical optics exact}

\author{T.\ G.\ Philbin}

\email{t.g.philbin@exeter.ac.uk}

\affiliation{Physics and Astronomy Department, University of Exeter,
Stocker Road, Exeter EX4 4QL, UK}

\begin{abstract}
Geometrical optics (GO) is widely used in studies of electromagnetic materials because of its ease of use compared to full-wave numerical simulations. Exact solutions for waves can, however, differ significantly from the GO approximation. In particular, effects that are ``perfect" for waves cannot usually be derived using GO. Here we give a method for designing materials in which GO is exact for some waves. This enables us to find interesting analytical solutions for  exact wave propagation in inhomogeneous media. Two examples of the technique are given: a material in which two point sources do not interfere, and a perfect isotropic cloak for waves from a point source. We also give the form of material response required for GO to be exact for all waves.
\end{abstract}

\maketitle

%%%%%%%%%%%%%%%%%%%%%%%%%%%%%%%%%%%%%%%%%%%
\section{Introduction}
The development of numerical solvers has enabled physicists and engineers to predict with great accuracy the propagation of electromagnetic waves in the most complex media, provided the effective medium picture holds. These numerical tools enable the optimisation of device designs in advance of their experimental implementation. It is doubtful, however, if numerics alone has ever revealed interesting effects in electromagnetic materials. The space of all possible inhomogeneous, anisotropic materials is simply too vast to explore numerically. When progress is made, the initial insight invariably comes from other theoretical sources, usually intuition supported by approximate analytical calculation. The need for guidance by analytical techniques is illustrated by the development of metamaterials~\cite{sarychev,cui,cai}, in which the search for applications and novel effects has been heavily influenced by transformation optics~\cite{leo}. One of the remarkable features of transformation optics is its ability to generate exact analytical results for wave propagation in inhomogeneous, anisotropic media. The tools of transformation optics deal with a very limited set of materials, however, and analytical results on the optics of inhomogeneous media have mostly been based on geometrical optics (GO). A great deal of information can be deduced using GO but the regime in which the GO approximation breaks down is also interesting. For example, the distinction between GO and exact wave optics has been central in understanding the limitations and possibilities of electromagnetic cloaking devices~\cite{leo06,pen06,leo08,per11}.

GO is not just ray tracing; it is {\em approximate} wave optics. The ray trajectories of GO are orthogonal to the wave fronts as they appear in the GO approximation, and the spacing between the GO wavefronts is obtained from the eikonal equation~\cite{born}. The exact wave fronts (in general) differ from those of GO and their derivation is usually vastly more difficult than ray tracing and solving the eikonal equation.  Working within the GO approximation allows design ideas to be developed far more easily, and if methods can be found to make GO solutions {\it exact} then designing interesting solutions for exact wave propagation becomes more feasible. The question of whether GO is exact is dependent on the optical medium and on the spatial form of the amplitude of the wave (see section~\ref{sec:GO}). Familiar examples of waves in homogeneous media for which GO is exact are plane waves and scalar spherical waves, but GO is not exact for simple wave forms such as vector spherical waves and (scalar or vector) cylindrical waves.  In this paper we describe a method for finding medium/wave combinations for which GO is exact. To simplify the discussion we restrict the analysis to scalar waves obeying the Helmholtz equation; a similar theory can be developed for vector waves. In section~\ref{sec:GO} we recall the GO approximation and section~\ref{sec:method} describes the method for deriving wave solutions for which GO is exact. Two examples of the method are presented in section~\ref{sec:ex}. Our method generates refractive-index profiles in which GO is exact for a particular wave; the type of material response required for GO to be exact for all waves is discussed in section~\ref{sec:res}.

%%%%%%%%%%%%%%%%%%%%%%%%%%%%%%%%%%%%%%%%%%
\section{Geometrical optics for scalar waves}  \label{sec:GO}
We consider monochromatic scalar waves in an isotropic, inhomogeneous medium, satisfying the Helmholtz equation
\begin{equation}  \label{helm}
\left[\bm{\nabla}^2+\frac{\omega^2}{c^2}n^2(\bm{r},\omega)\right]\psi(\bm{r},\omega)=0.
\end{equation}
A similar analysis to what follows can be performed for vector waves using the derivation of the GO approximation from Maxwell's equations~\cite{born}. If the frequency-domain, complex wave $\psi(\bm{r},\omega)$ is written in terms of its amplitude and phase,
\begin{equation}  \label{psiRS}
\psi(\bm{r},\omega)=R(\bm{r}){\rm e}^{{\rm i}S(\bm{r})},
\end{equation}
where $R(\bm{r})$ and $S(\bm{r})$ are real, then (\ref{helm}) gives the two real equations
\begin{gather} 
(\bm{\nabla}S)^2-\frac{\bm{\nabla}^2R}{R}-\frac{\omega^2}{c^2}n^2(\bm{r},\omega)=0, \label{SR1} \\
\bm{\nabla\cdot}(R^2\bm{\nabla}S)=0.  \label{SR2}  
\end{gather}
The GO approximation corresponds to neglecting the second term in (\ref{SR1}), which then reduces to the eikonal equation of GO~\cite{born}:
\begin{equation}  \label{GO}
(\bm{\nabla}S)^2-\frac{\omega^2}{c^2}n^2(\bm{r},\omega)=0.
\end{equation}
Comparing (\ref{SR1}) with (\ref{GO}), we see that GO gives an approximate solution for the phase $S(\bm{r})$ of the wave. The ray trajectories of GO lie on the gradients $\bm{\nabla}S$ of this approximate phase and are therefore not (in general) orthogonal to the exact phase fronts of the wave. Even in those cases where the GO ray trajectories are orthogonal to the exact phase fronts, the GO approximate phase $S(\bm{r})$ may not give the exact phase accumulation along the ray (cylindrical waves provide a simple example of this). The exact phase $S(\bm{r})$ is coupled to the amplitude $R(\bm{r})$ via the non-linear equations (\ref{SR1})  and (\ref{SR2}), which are very difficult to solve analytically except in a few simple cases. It is the decoupling of the phase from the amplitude in (\ref{GO}) compared to  (\ref{SR1}) that makes GO more tractable mathematically; once the GO phase is found from (\ref{GO}) the amplitude is separately determined by (\ref{SR2}). 

It is curious that the optics community has not given a name to the term $\bm{\nabla}^2R/R$ in (\ref{SR1}), despite the fact that this term is responsible for the difference between exact wave optics and GO. In quantum mechanics, on the other hand, there is a name for this term---it is called the {\em quantum potential}~\cite{bohm,holland,durr}. The Helmholtz equation (\ref{helm}) is the time-independent Schr\"{o}dinger equation and the quantum potential $\bm{\nabla}^2R/R$ in (\ref{SR1}) is the main reason why the predictions of quantum mechanics differ from those of classical mechanics (the other reason is the requirement  for $\psi$ to be single valued)~\cite{bohm,holland,durr}. The term $\bm{\nabla}^2R/R$ is just as significant in optics as it is in mechanics; given the lack of an optical designation for this term we will refer to it here as the quantum potential.

%%%%%%%%%%%%%%%%%%%%%%%%%%%%%%%%%%%%%%%%%
\section{Method}  \label{sec:method}
If the quantum potential vanishes, i.e.\ if the amplitude $R(\bm{r})$ of the wave satisfies
\begin{equation}  \label{qpzero}
\frac{\bm{\nabla}^2R}{R}=0,
\end{equation}
then GO is exact and the solution of (\ref{GO}) gives the exact phase $S(\bm{r})$ and therefore the exact wave fronts. The amplitude $R(\bm{r})$ can then be found by solving (\ref{SR2}). Equation (\ref{qpzero}) holds for plane waves and spherical waves, for example, and GO is therefore exact for these waves. The relation (\ref{qpzero}) is not restricted to $r\neq0$ for a spherical wave $e^{ikr}/r$: the amplitude $R(\bm{r})$ for this wave goes as $r^{-1}$ and so $\bm{\nabla}^2R/R\propto r\bm{\nabla}^2r^{-1}\propto r\delta(r)=0$, for all $r$. But the spherical wave has a point source at $r=0$ so that the Helmholtz equation (\ref{helm}) is not valid at $r=0$ (a delta-function source term is required on the right-hand side). Equation (\ref{qpzero}) does not hold for cylindrical waves so GO is not exact in this case. A cylindrical wave is proportional to a Hankel function whereas the GO solution is proportional to $e^{{\rm i}kr}/\sqrt{r}$, the asymptotic limit of the Hankel function. As noted above, the cylindrical wave provides an example where the GO rays are orthogonal to the exact phase fronts but the GO phase differs from the exact phase.

Note that if
\begin{equation}  \label{StoR}
S(\bm{r})\propto \frac{1}{R(\bm{r})}
\end{equation}
then the exact wave equation (\ref{SR2}) becomes
\begin{equation} \label{lap}
\bm{\nabla}^2R=0,
\end{equation}
which gives (\ref{qpzero}) (barring any zeros in amplitude), and this in turn implies (\ref{GO}). We can therefore generate a wave solution in a refractive index profile for which GO is exact as follows. Choose a solution of the Laplace equation (\ref{lap}) for the amplitude $R(\bm{r})$ and choose the phase $S(\bm{r})$ to be inversely proportional to $R(\bm{r})$. Then one part of the exact wave equation, namely (\ref{SR2}), is automatically satisfied and the second part, equation (\ref{SR1}), reduces to the eikonal equation (\ref{GO}) which can be solved for $n(\bm{r},\omega)$ since $S(\bm{r})$ is known. The derived wave is by design exactly described by GO; for other waves in the derived index profile, GO will not be exact.

%%%%%%%%%%%%%%%%%%%%%%%%%%%%%%%%%%%%%%%%%%%%
\section{Examples}    \label{sec:ex}

\subsection{Waves without interference}
\begin{figure}[!htbp]
\centerline{\includegraphics[width=9cm]{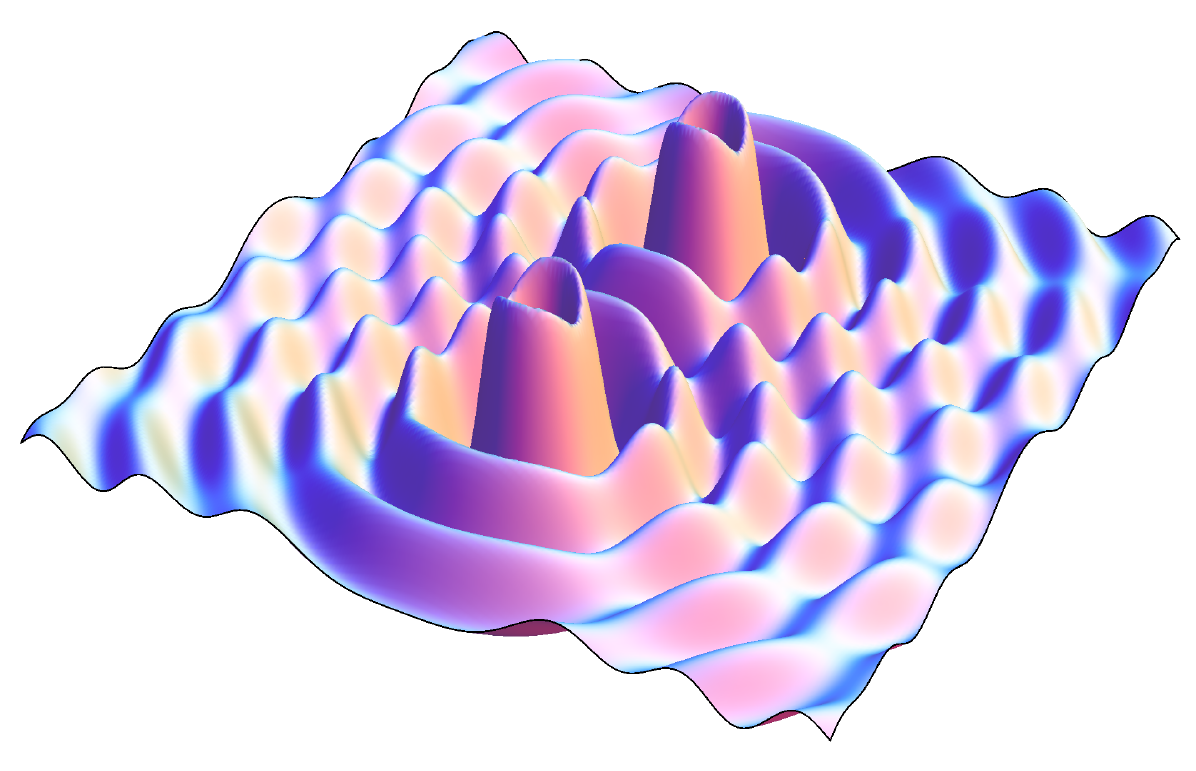}}
\caption{Waves from two point sources in vacuum. The plot shows the amplitude of the wave in a two-dimensional slice through the point sources.}
\label{fig:pointsvac}
\end{figure}

\begin{figure}[!htbp]
\centerline{\includegraphics[width=9cm]{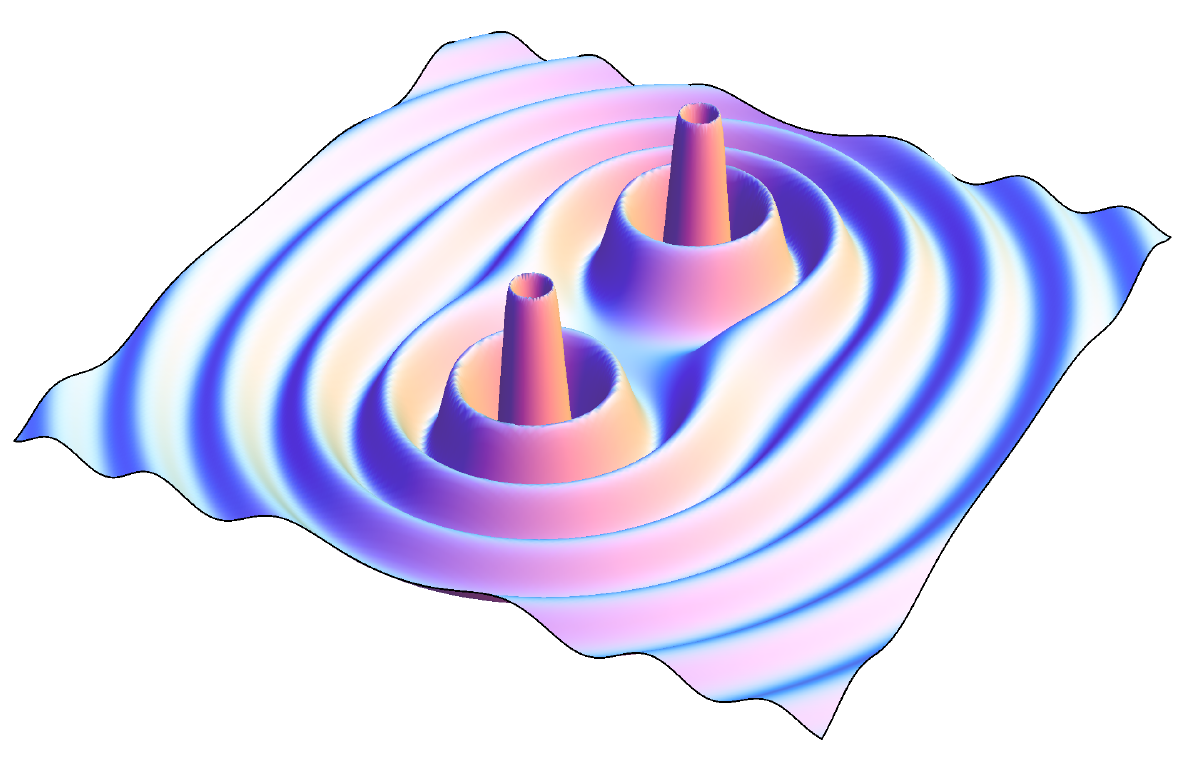}}
\centerline{\includegraphics[width=8.5cm]{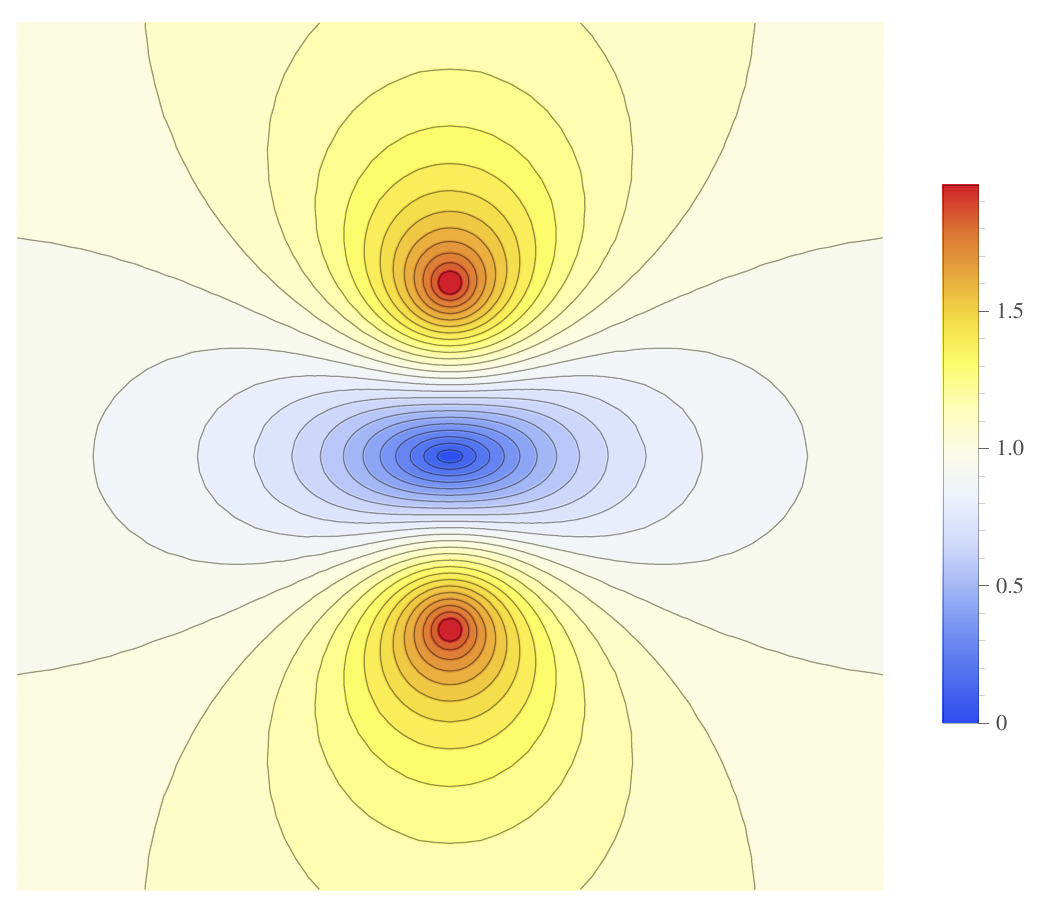}}
\caption{Wave from two point sources (top) in an inhomogeneous refractive index profile (bottom), both shown for a 2D slice through the position of the sources. The wave shows no interference or scattering, regardless of whether the index changes significantly over a wavelength. The refractive index ranges from $2$ at the source positions to $0$ at one point midway between the sources.}
\label{fig:wwi}
\end{figure}

Consider two point sources in vacuum, located at positions $(0,Y,0)$ and $(0,-Y,0)$. These sources separately produce spherical waves with amplitudes
\begin{align}
R_1(\bm{r})&=-\frac{1}{4\pi\sqrt{x^2+(y-Y)^2+z^2}},   \label{R1}  \\
R_2(\bm{r})&=-\frac{1}{4\pi\sqrt{x^2+(y+Y)^2+z^2}},   \label{R2}
\end{align}
respectively. The two spherical waves interfere as shown in Fig.~\ref{fig:pointsvac}. Applying the technique described in the previous section we choose the amplitude $R(\bm{r})$ and phase $S(\bm{r})$ of a new wave to be
\begin{equation}   \label{nowave}
R(\bm{r})=\frac{1}{2}\left(R_1(\bm{r})+R_2(\bm{r})\right), \quad S(\bm{r})=-\frac{\omega}{4\pi c R(\bm{r})}.
\end{equation}
%%%%%%%%%%%%%%%%%%%%%%%%%%%%%%%%%%%%%%
%%%%%%%%%%%%%%%%%%%%%%%%%%%%%%%%%%%%%%
The amplitude $R(\bm{r})$ satisfies Laplace's equation because $R_1(\bm{r})$ and $R_2(\bm{r})$ are solutions of this equation; the wave (\ref{nowave}) therefore meets the conditions, described in the previous section, for GO to be exact. The refractive index in which this wave propagates is given by substituting the phase $S(\bm{r})$ into the eikonal equation (\ref{GO}). As the amplitude $R(\bm{r})$ in (\ref{nowave}) is chosen to be the average of the two point-source amplitudes, there is no interference in the resulting wave $R(\bm{r})e^{iS(\bm{r})}$. The wave propagation and refractive index profile in a 2D slice through the sources are shown in Fig.~\ref{fig:wwi}. The lack of any interference or scattering off the index profile is here an exact result, independent of the gradient of the refractive index compared to the wave-vector. It is relatively easy to design index profiles in which the GO ray trajectories for two point sources will be qualitatively the same as in the index profile in Fig.~\ref{fig:wwi}, but such profiles will give scattering of the wave off the inhomogeneous material when the GO approximation is invalid. In contrast, the wave solution in Fig.~\ref{fig:wwi} is exact in the index profile shown. The refractive index ranges from $2$ at the sources to zero at one point midway between them; the index approaches $1$ at large distances from the sources. As the zero in the index occurs at an isolated point, it can be removed by the transmutation procedure of transformation optics~\cite{tyc08,leo} at the cost of introducing some anisotropy in the transmuted region.

%%%%%%%%%%%%%%%%%%%%%%%%%%%%%%%%%%%%%%
%%%%%%%%%%%%%%%%%%%%%%%%%%%%%%%%%%%%%%

%%%%%%%%%%%%%%%%%%%%%%%%%%%%%%%%%%%%%%%%%
\subsection{Cloaking of waves with an isotropic material}
\begin{figure}[!htbp]
\centerline{\includegraphics[width=8cm]{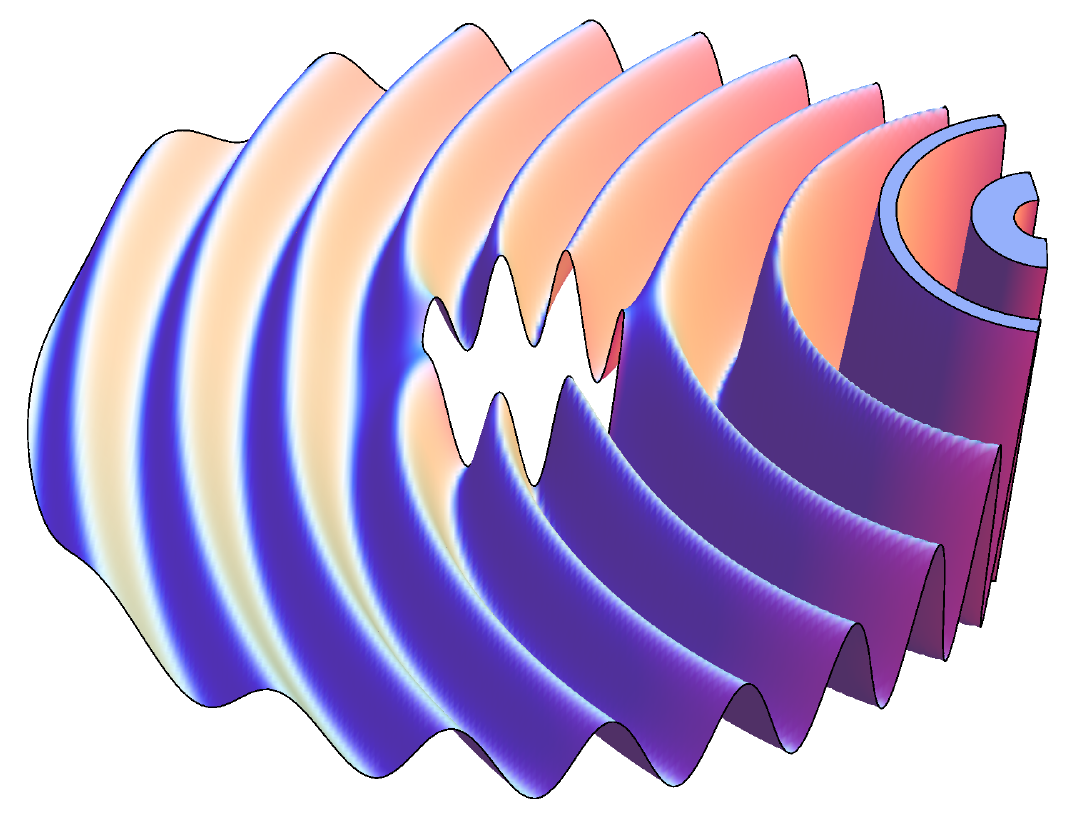}}
\centerline{\includegraphics[width=8.5cm]{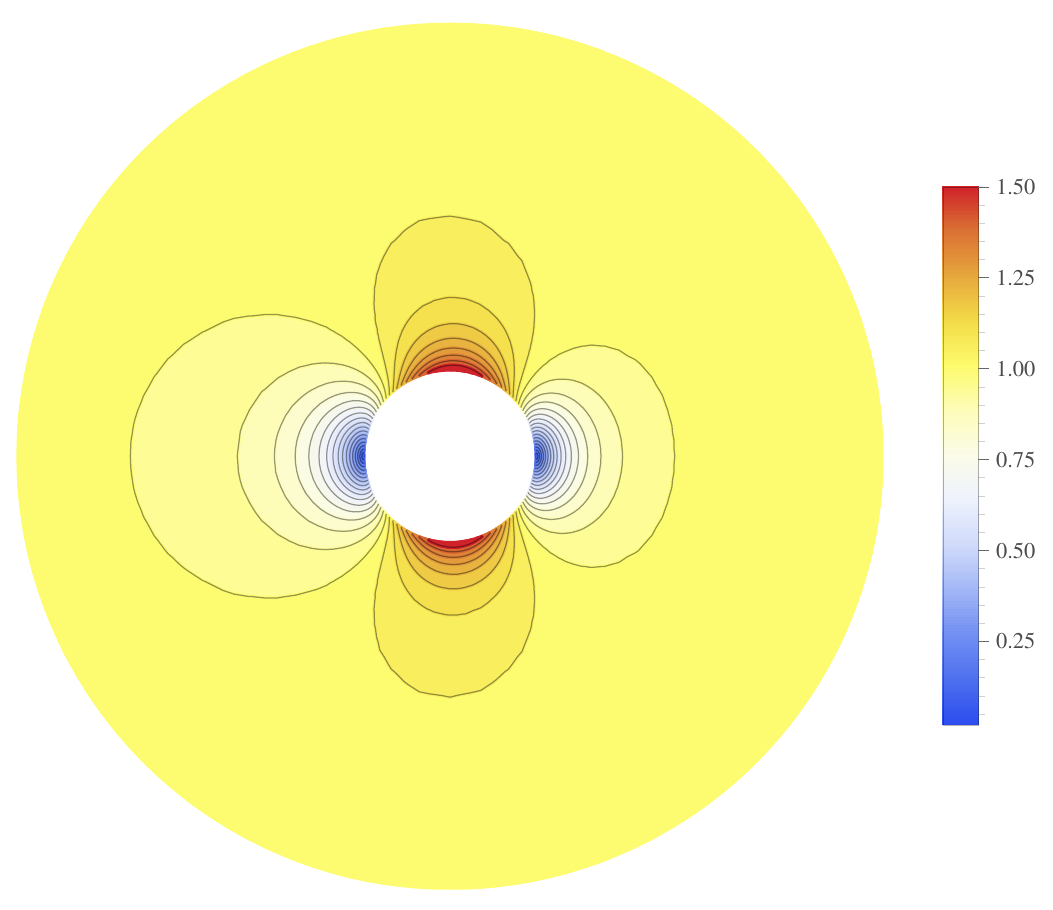}}
\caption{Wave from a point source (top) in an inhomogeneous index profile (bottom). The plots show the wave and index profile in the $xz-$ or $yz$-plane. The wave is guided around a cloaked spherical region and at large distances is indistinguishable from the wave produced by a point source in vacuum.}
\label{fig:cloak}
\end{figure}

Consider a point charge located at $(0,0,z_0)$, outside a zero-permittivity ball of radius $b$ centred on the origin. The electric-field lines from the charge are guided around the ball in much the same way as rays are guided around a cloaked region. Guiding of magnetic-field lines around zero-permeability (superconducting) objects is more familiar, but here we wish to have a point source for the field so we consider an electrostatic example of such field exclusion. Using the standard boundary-value methods of electrostatics~\cite{jac}, it is straightforward to show that the electric potential outside the ball is (for unit charge and with $\varepsilon_0=1$)
\begin{align}
\phi(\bm{r})=&\frac{1}{4\pi}\left[\frac{1}{\sqrt{x^2+y^2+(z-z0)^2}} \right.  \nonumber \\
& \quad\ \    \left. + \sum_{l=0}^\infty\frac{lb^{2l+1}}{(l+1)(z_0r)^{l+1}}P_l\left(\frac{z}{r}\right) \right], \quad r>b,
\end{align}
where $r=\sqrt{x^2+y^2+z^2}$. We choose the amplitude $R(\bm{r})$ and phase $S(\bm{r})$ of our wave to be
\begin{equation} \label{cloakwave}
R(\bm{r})=-\phi(\bm{r}), \quad S(\bm{r})=-\frac{\omega}{4\pi c R(\bm{r})}.
\end{equation}
Since the potential $\phi(\bm{r})$ satisfies Laplace's equation, the wave (\ref{cloakwave}) meets the conditions of section~\ref{sec:method} for GO to be exact. The wave propagation and the refractive-index profile, in a 2D slice through the source and the centre of the cloaked region, are shown in Fig.~\ref{fig:cloak}. The wave is guided around the spherical region $r<b$ which is cloaked. Far from the cloaked region the wave approaches that of a point source in vacuum. The maximum refractive-index value is $1.5$, at two isolated points on the boundary of the cloaked region lying on a line orthogonal to the direction to the source. The minimum index value is $0$, at two isolated points on the boundary of the cloaked region lying on a line through the centre of the cloaked region and the source. The two zeros of the refractive index can be removed by transmutation~\cite{tyc08,leo}. Note that this index profile only cloaks the region $r<b$ when a point source is placed at $(0,0,z_0)$; waves from other sources in this index profile will not be cloaked. This is in line with a general theorem that shows the impossibility of cloaking waves from all directions using an isotropic material~\cite{nac88}.

%%%%%%%%%%%%%%%%%%%%%%%%%%%%%%%%%%%%%%
%%%%%%%%%%%%%%%%%%%%%%%%%%%%%%%%%%%%%%

As in the previous example, it is not difficult to design index profiles in which the GO rays have the same qualitative behaviour as in the index profile of Fig.~\ref{fig:cloak}. Guiding of GO rays around some region is in fact the only feasible method for broadband omnidirectional electromagnetic cloaks, since perfect omnidirectional cloaking of waves is not strictly possible~\cite{leo06,pen06,leo08,per11}. The example derived here is different because GO is exact: there is no scattering of the wave off the inhomogeneous index profile, even if the wavelength is such that the gradient of the refractive index is comparable to the wave-vector. There is however a matching issue at the boundary of the cloaked region. In the omnidirectional perfect wave cloak~\cite{pen06}, the cloaked region is electromagnetically cut off from the exterior by a surface of zero refractive index (this surface of zero index cannot be transmuted away~\cite{tyc08,leo}, rendering the perfect omnidirectional cloak impossible in practice). In the example derived here there is no such cut off, which means the wave can evanescently probe the cloaked region. This may degrade the cloaking effect compared to the ideal perfect cloaking in Fig.~\ref{fig:cloak}.

%%%%%%%%%%%%%%%%%%%%%%%%%%%%%%%%%%%%%%%%%
\section{Material response for exact geometrical optics}   \label{sec:res}
The method of section~\ref{sec:method} constructs a particular wave solution in a refractive-index profile such that GO is exact for the wave. An obvious question is whether there exists a material in which GO is exact for {\em all} waves. In terms of a refractive-index profile the answer to this question is negative, but if we allow for a material response not describable solely by a refractive index, then there exists a wave equation for which GO is exact for all waves. Consider the following modification of the Helmholtz equation:
\begin{equation}  \label{helmmod}
\left[\bm{\nabla}^2+\frac{\omega^2}{c^2}n^2(\bm{r},\omega)\right]\psi-\frac{\bm{\nabla}^2|\psi|}{|\psi|}\,\psi=0.
\end{equation}
If we again write the complex wave $\psi(\bm{r},\omega)$ in terms of its amplitude and phase, as in (\ref{psiRS}), then (\ref{helmmod}) is equivalent to the two real equations
\begin{equation} 
(\bm{\nabla}S)^2-\frac{\omega^2}{c^2}n^2(\bm{r},\omega)=0, \qquad
\bm{\nabla\cdot}(R^2\bm{\nabla}S)=0,
\end{equation}
which are exactly the equations of GO. The material response described by the last term in (\ref{helmmod}) is very unusual: it is not a nonlinear response because it scales linearly with the amplitude of the wave. The factor $\bm{\nabla}^2|\psi|/|\psi|$ describes an effective refractive index that depends on the factional spatial variation of the wave amplitude. It would be of great interest if a physical system could be found in which (\ref{helmmod}) describes wave propagation---since GO would be exact for such waves, they would exhibit no interference effects. For example, a two-slit interference experiment would show an intensity pattern with no fringes; the classic demonstration of the wave nature of light would in this case fail to show any wave behaviour. 

 As noted in section~\ref{sec:GO}, the Helmholtz equation is the time-independent Schr\"{o}dinger equation and we have seen that modifying the latter to read (\ref{helmmod}) corresponds to zero quantum potential for all waves. Vanishing quantum potential is the requirement for the predictions of quantum mechanics to be exactly classical~\cite{bohm,holland,durr}. It was noted long ago that addition of a term $-\psi\bm{\nabla}^2|\psi|/|\psi|$ to the Schr\"{o}dinger equation removes all quantum effects~\cite{sch62,holland}. As with the significance of the quantum potential generally, these insights are not often translated into optical language where they illuminate the relation between exact wave optics and GO.

%%%%%%%%%%%%%%%%%%%%%%%%%%%%%%%%%%%%%%%%%%
\section{Conclusions}
 In situations where GO is exact the problem of wave propagation simplifies enormously and exact analytical solutions are greatly facilitated. We have developed a method for generating wave solutions in inhomogeneous refractive-index profiles for which GO is exact. The method allows the exploration of exact wave propagation in inhomogeneous media without resorting to full-wave numerical simulations. Two examples of the technique were given (i) an index profile in which two point sources at specified positions do not interfere, and (ii) an index profile that cloaks the wave from a point source at one position. We have also pointed out the kind of material response required for GO to be exact for all waves.

 %%%%%%%%%%%%%%%%%%%%%%%%%%%%%%%%%%%%%%%%%%%
\section*{Acknowledgements}
EPSRC provided financial support under Program Grant EP/I034548/1.

\end{document}